# Acoustic-Friction Networks and the Evolution of Precursory Rupture Fronts in Laboratory Earthquakes

H.O. Ghaffari[*], and R.P. Young

**Affiliations**
Department of Civil Engineering and Lassonde Institute, University of Toronto, 170 College Street, Toronto, ON M5S3E3 Canada

The evolution of shear rupture fronts in laboratory earthquakes are analysed with the corresponding functional networks, constructed over acoustic emission friction-patterns. We show that the mesoscopic characteristics of functional networks carry the characteristic time for each phase of the rupture evolution. The classified rupture fronts in network states–obtained from a saw-cut fault and natural faulted Westerly granite - show a clear separation into three main groups, indicating different states of rupture fronts. With respect to the scaling of local ruptures' durations with the networks' parameters, we show that the gap in the classified fronts could be related to the possibility of a separation between slow and regular fronts.

**Introduction-** Recently, a series of laboratory earthquake/friction experiments have reported observing complex slip-front evolution in terms of the speed and the general configurations of the ruptures [1-12]. Although some aspects of the laboratory (or field observations) had been predicted through numerical experiments [9-21], the new evidence necessitates a more precise analysis of the results and observations. In considering the complexity of rupture fronts, the super-shear fronts, slow fronts and mixture ruptures (i.e., transitions in rupture modes) as well as precursory events preceding the main stick-slip events have been of interest. Study of precursor events in laboratory earthquakes -as the arrested or potential of transition rupture- revealed a new type of slow dynamic of the frictional interfaces. Ohnaka and Fineberg [1, 11] reported a slow front evolution during frictional slipping, which was followed by several other numerical and experimental works [3, 5, 9, 11, 13, 17-18, 28]. In particular, it was proposed that the occurrence of slow ruptures in laboratory scale accompanies weak acoustic amplitude signatures in radiated acoustics [1, 2]. A recent robust modeling highlighted the intrinsic nature of slow ruptures that resulted from rate-and-state friction laws, which led to the prediction of a new velocity scale as the minimum rate of slow rupture propagation [17-18]. The observed slow ruptures can be compared with slow earthquakes as a new topic while their mechanisms are not completely understood [18, 22-28]. Also, analyzing the collection of large-scale earthquakes regarding source duration and the final displacement of events has shown a clear gap between regular earthquakes and silent earthquakes, representing different classes of slip front propagation [22, 27].

From another perspective, study of the evolution of frictional interfaces with respect to the passage of Sub-Rayleigh precursory rupture fronts on poly methyl methacrylate (PMMA) blocks showed several phases of the evolution from the arrested rupture fronts [6]. Fast approaching rupture tip leading to failure of asperity and large deformation is proceed with the fast slip while a time characteristic controls the distinguished phases. In this research, we show that functional acoustic networks also show the similar evolutionary phases while we investigate the rupture fronts from Westerly Gran-ite. We speculate that the dynamic configurations of multi-channel acoustic emission waveforms in appropriate network-parameter spaces show different possible classes of rupture regimes, encoded in the corresponding constructed functional networks from the recorded signals. Also, we provide solid observations of the gap between regular and slow fronts while the rupture fronts are analyzed in terms of their local-source duration and network-local energy flow index. Our results clearly indicate that precursory events from dry frictional interface follow a universal trend in corresponding network states, separating their features from random (null) networks.

**Results -** Our data set includes the recorded discrete and continuous waveforms (i.e., acoustic emissions-AEs) using 16 piezoelectric transducers from a saw-cut sample of Westerly granite (LabEQ1), under triaxial loading [31-32]. The saw cut was at a 60 degree angle and polished with silicon carbide 220 grit. Each triggered event had the duration of 204.8 µs (recorded at 5 MHz), while the three main stick-slip events occurred with shear stress drops of 54, 56 and 79 MPa. The experiment was servo-controlled using an axial strain rate of $5 \times 10^{-6} s^{-1}$. The confining stress was maintained at 150 MPa for three reported main stick-slip events, producing 109 located- rupture fronts events. The second data set (LabEQ2) are the results of the two main cycles of loading –unloading (stick-slip) of Westerly granite on a preexisting natural fault by loading at constant confining pressure. Natural rough fault was created using a triaxial loading system at constant confining pressure of 50 MPa and with acoustic emission feedback control. The two main phases of stick-slip were accomplished in 200 MPa and 150 MPa confinement pressure, respectively. The driving strain rate was identical to the LabEQ1. We report an analysis of over 8000 recorded rupture fronts from two main stick-slip events. Using the aforementioned data set, we construct the corresponding functional networks (see the methods part) and study their properties such as betweenness centrality (B.C), maximum modularity (Q), with-in module degree (Z), maximum eigenvalue of Laplacian of connectivity matrix and inverse participation ratio (P). The evolutions of corresponding functional friction networks are rationalized in terms of their local and mean attributes (i.e., spatial and temporal mean).

Starting with LabEq.1data set, we show that most of the recorded events exhibit a universal trend of the evolution in terms of their modularity trend (Fig1a, b). Considering the modularity as the mean response of the whole nodes (corresponding to stations) and the rough collapse of the data (Fig 1b), one can recognize the 4 main evolutionary phases during 204 $\mu s$ time-window (where the calculated maximum modularity value varies between 0.001 and 0.49). Phase I is quantified with the fast drop of the modularity and the approaching to the minimum modularity. We assign this phase to the approaching and large deformation of the asperities (or asperity). In other words, the mean dramatic deformation of a "node" is encoded in fast drop of the modularity while mean local energy flow index ( $< B.C >_x$ -Eq.2 Methods part) is also reaching to minimum value. Our investigation on ~50 regular events showed this phase has duration between $\sim 8 \pm 4$ $\mu s$ (we will show this interval is longer in slow deformation). The magnitude of this drop is scaled with the maximum recorded of voltage from the employed piezoelectric sensors. Larger deformation and crushing asperities with significant damages is encoded in big amount of the drop in the modularity value. A rapid modularity growth phase is the characteristics of the phase 2 with the duration of $\sim 25 \pm 7$ $\mu s$ . Possible explanation of the fast rising time is to consider the occurrence of the rapid slip of broken asperities while after a nearly constant timescale a gradual drop is observed (Phase III). The next following phase (IV) is generally decaying stage much slower than previous phase, with a small range of fluctuations in $Q$. The duration of this phase is the longest one in comparison with the previous phases. The decay rate of this phase is significantly higher for Lab.Eq2, indicating the role of heterogeneity of rough-faults in suppressing slip signals (Fig.S4, S5, S6). Regardless of the real physical meaning of Q ,the similarity of evolutionary phases of the network-parameter space of the precursory events to the reported slip profiles from glassy materials (such as PMMA) [6], shows that the fracture inducing weakening is general scenario in frictional interfaces. However, the mechanism controlling the evolutionary phases for granite samples cannot be explained with plastic deformation, as the production of thermal weakening [6]. The alternative approach is to consider a substantial cracking or crushing the asperities leading to wearing process.

Considering the spatio-temporal mean value of betweenness centrality over LabEq.1, we find out the three distinguished classes of rupture fronts (Fig.2c). The main stick-slip events are mapped into R1 zone which indicates the high amplitude activities of acoustic emission waveforms. This zone generally accompanies the visible records of displacements, mostly recorded in stick-slip periods. The most recorded events are encoded in R2, where we classify them as "regular" events. A few of the precursory rupture fronts allocate R3 class. We found that unusually long localizations of eigenvectors ( $\langle P \rangle_x$ -Eq.2 in Methods part) regarding regular duration (such as R2 ruptures) with high maximum eigenvalues are the highlighted features of R3 (Fig.3.a and d). The long term of localization is nearly 2 to 5 times the regular localization period (i.e., R2 ruptures (Fig.3.c)). This picture is compatible with the longer duration of phase I . We carefully tested the amplitudes of acoustic waveforms from the R3 events which indicated a weak-long activity of the signals in most of the stations. We observe these features at all locations traversed by all well-located triggered rupture fronts as have been shown in Fig.2.b. Another feature which distinguishes R3 events from R2 fronts is shown in Fig.3a: a unique gap between R2 and R3 in $\overline{\log < B.C >} - \overline{\lambda}_{max}$ parameter space. All evidences hint that the R3 fronts are slow ruptures with distinguished separation from regular ruptures, as it has been indicated in natural large slow earthquakes [22, 27].

Matching the far-field driving stress field with the corresponding individual arrested rupture fronts ,triggered at the same time ,reveals a nearly logarithmic decay decay of the temporal mean of the modularity versus shear stress per each cycle of loading in LabEq1(Fig.3.b): $\tau_s \simeq -0.33 \times 10^3 \log \overline{Q}$ in which $\overline{Q}$ is the temporal average over all phases. In other words, approaching to main stick-slip event is accompanied with a lower modularity index, indicting possible longer phase I (or significant amount of the drop from the rest state) or decreasing the peak value of the maximum modularity in phase I (also see Fig.S7). This shows how micro-ruptures, resulted in the recorded acoustic emission waveforms evolve regarding the uniform shear stress. Assuming the validation of the Amontons-Coulomb law $(\tau_S = \mu_S \sigma_N)$ in the scale of the arrested fronts, we reach to: $\mu_S \sim \log(\overline{Q}^{-\xi})$ for a constant normal loading, indicating that local static friction is possibly correlated with the details of the rupturing and slipping of the local fronts, comparable with the recent experimental results [9].

For all collapsed events in $\overline{\log < B.C >} - \overline{\lambda}_{max}$, we can assign a simple power law as well as : $\overline{\lambda}_{max} \sim \overline{\log < B.C >}^{-\beta}$ which also holds for LabEq.2 data set (Fig.4a). In contrast to LabEq1, events from the natural-rough fault does not show the separation and clear gap of R3 from R2 while transition to R2 is observed in terms of the decrease in the critical exponent of the scaling relation (Fig.4a-inset). This transition also can be followed in terms of the density of the events in R3 versus R2 (Fig.4b). While for Lab.Eq1, most of R3 fronts occur in initial loading steps, the temporal distribution of R3 in Lab.Eq2 is not limited to a specific time step. Investigation on the recurrence times of precursory ruptures in each class shows that the observation of the density changes indeed is mapped into the waiting times (Fig.S10). Considering Lab.Eq2 ,most of the events in R2 zone show waiting times < 200ms while events in R3 point out a combination of <200ms to 2000ms. In ruptures with the signature of significant release of energy (R1 class), the mean of the waiting times are around ~2000ms. The main implication of three main classes of recurrence times probably could be a point of more investigation on the role of precursory events on healing or aging phenomena on frictional interfaces. Events with longer waiting time allocate a certain class of the introduced parameter spaces (i.e., R1), while shorter recurrence time indicates relatively rupture fronts with weaker amplitude, possibly small stress drop per each front.

**Discussion-** The applications of complex networks on multi-stationary accused laboratory earthquakes regarding precursory rupture fronts were addressed in this research. We recognized the main evolutionary phases of each rupture front, imprinting approaching rupture front, crashing or failure of asperity (nodes in networks) and slipping phases. This finding provides a general universal mechanism for fracture weakening mechanism for precursor rupture fronts. The recorded signals from rough-frictional interface show a faster decay of slip-phase rather than smooth-faults while generally ruptures with higher maximum modularity decay faster. The possible scaling of average temporal modularity with the shear stress was observed in smooth fault, indicating the dependency of local static friction to fronts' phases, confirming recent experimental results on PMMA. Furthermore, the classification of the recorded rupture fronts using the mesoscopic (*i.e.,* modular features) and transport characteristics of friction networks revealed three main rupture fronts. The analyzed events from smooth and rough faults portrayed a universal trend of ruptures' evolution where the recorded weak acoustic waveforms were encoded in high modularity index within the separated distinct clusters with long-duration in rupturing phase (i.e., local source duration). The separated cluster with scatter events was related to observation of slow ruptures in natural earthquakes. Obviously, with respect to clipping amplified acoustic waveforms for events with high release of energy (see figure S9), we cannot investigate possible slow events in large amount of seismic moments. Interestingly, analysis over 8000 rupture fronts showed that the functional friction networks are not random networks and imprint a unique signature in different network parameter spaces. The recurrence time of each class in the introduced scalar parameter space was related the maximum modularity of the events in that class : generally high energy precursory events have longer waiting times while very weak rupture fronts may have abnormally short or long waiting time. The application of the presented methods on natural earthquakes and comparing their features with the laboratory weak-rupture fronts will be our further work.

**Methods** – We propose the following methods to characterize interface evolutions within the available data set. To set up a non-directed network , we use the meta-time series method [33] in which the multi-channel (station), simultaneously-recorded time series is mapped onto a proper network. The method is used over the recorded time-series from acoustic transducers. A similar method has been used on real-time contact areas and apertures of frictional interfaces [29-30]. We start with the normalization of waveforms in each station and then the division of $N$- recorded time series with the length of $T$ into $m$ segments. The $j$th segment from $i$th time series ($1 \leq i \leq N$) denoted by $x^{i,j}(t)$ is compared with $x^{k,j}(t)$ to make an edge among the aforementioned segments. If the $j$th segment of the $i$th and $k$th time series are "close" enough to each other, we set $a_{ik}(j)=1$ otherwise $a_{ik}(j)=0$ in which $a_{ik}(j)$ is the component of the connectivity matrix. We use a "closeness" metric: $d(x^{i,j}(t), x^{k,j}(t)) = \sum_t \left\| x^{i,j}(t) - x^{k,j}(t) \right\|$. To precisely analyze a time series with the aforementioned methods and reduce possible errors due to the limited number of stations, we set $m=1$ (equal to each recorded point). We also increased the size of the adjacency matrix with simple interpolation of $d$ using cubic spline interpolation. The increasing of number of nodes generally did not change the presented results and just increased the quality of visualization of the results. Then for the acoustic-friction networks, the numbers of nodes are 50 and we keep the number of nodes as the constant value.

To proceed, we use several characteristics of networks. Each node is characterized by its degree $k_i$ and the clustering coefficient. The clustering coefficient (as a fraction of triangles) is $C_i$ defined as $C_i = \frac{2T_i}{k_i(k_i-1)}$ where $T_i$ is the number of links among the neighbors of node $i$ and $k_i$ is the number of links. For a given network with $N$ nodes, the degree of the node and Laplacian of the connectivity matrix are defined by $k_i = \sum_{j=1}^{N} a_{ij}; L_{ij} = a_{ij} - k_i \delta_{ij}$ where $k_i, a_{ij}, L_{ij}$ are the degree of $i$ th node, elements of a symmetric adjacency matrix, and the network Laplacian matrix, respectively. The eigenvalues $\Lambda_\alpha$ are given by $\sum_{j=1}^{N} L_{ij} \phi_j^{(\alpha)} = \Lambda_\alpha \phi_i^{(\alpha)}$, in which $\phi_i^{(\alpha)}$ is the $i$ th eigenvector of the Laplacian matrix ($\alpha = 1,...,N$). With this definition, all eigenvalues are non-positive values. A scalar measure of the localization degree of a vector is called the inverse participation index. The inverse participation ratio as a criterion of the localization of eigenvectors is defined by [33]:

$$P(\phi^\alpha) = \frac{\sum_i (\phi_i^\alpha)^4}{(\sum_i (\phi_i^\alpha)^2)^2}; \alpha = 1,...,N \tag{1}$$

The maximum value of $P$ shows that the vector has only one non-zero component. A higher value of $P$ corresponds with a more localized vector. We also address the role of betweenness centrality (B.C) of a node as the measure of "load" [35]:

$$B.C_i = \frac{1}{(N-1)(N-2)} \sum_{\substack{h,j \\ h \neq j, h \neq i, j \neq i}}^{N} \frac{\rho_{hj}^{(i)}}{\rho_{hj}} \tag{2}$$

in which $\rho_{hj}$ is the number of the shortest path between $h$ and $j$ ,and $\rho_{hj}^{(i)}$ is the number of the shortest path between $h$ and $j$ that passes $i$. We also use the networks' modularity characteristics. In particular, based on the role of a node in the network modules, each node is assigned to its within-module degree ($Z$) and its participation coefficient ($P$). High values of $Z$ indicate how well-connected a node is to other nodes in the same module, and P is a measure of well-distribution of the node's links among different modules [36]. The modularity $M$ (i.e., objective function) is defined as [37-38]:

$$M = \sum_{s=1}^{N_M} \frac{l_s}{L} - \left( \frac{d_s}{2L} \right)^2 ], \tag{3}$$

in which $N_M$ is the number of modules (clusters) , $L = \frac{1}{2} \sum_i^N k_i$ , $l_s$ is the number of links in module and $d_s = \sum_i k_i^s$ (the sum of nodes degrees in module s). Using an optimization algorithm, the cluster with maximum modularity (Q) is detected. The correlation of a node with the degree of neighbouring nodes is defined as assortatitive mixing index [35]:

$$r_k = \frac{<j_i k_i> - <k_i>^2}{<k_i^2> - <k_i>^2} \qquad (4)$$

where it shows the Pearson correlation coefficient between degrees $(j_i, k_i)$ and $<\bullet>$ denotes the average over the number of links in the network. High assortativity indicates the attraction of rich nodes to each other (i.e., hubs) and negative value of $r$ presents disassortative attribute of nodes where "poor" nodes are attracted to hubs.

### Acknowledgements

We would like to acknowledge and thank Dr. B.D. Thompson (Mine Design Engineering, Kingston, Canada), Dr. D. Lockner (USGS, Menlo-Park, USA), Prof. S. Nielsen (INGV, Italy), Dr. A. Schubnel (Laboratoire de Géologie de l'Ecole normale supérieure, France), Prof. J. Fineberg and Dr. O.Ben-David (The Racah Institute of Physics, Hebrew University of Jerusalem) for providing the employed data set in this work. The first author would like to express appreciation to Prof. K. Xia (University of Toronto) and Dr. A. Schubnel for their comments and points on this manuscript.

### Contributions

All authors contributed equally to the work presented in this paper including ideas, manuscript preparation and analysis.

### Competing financial interests

The authors declare no competing financial interests.

### Corresponding author

Correspondence to: H .O. Ghaffari (h.o.ghaffari@gmail.com)

**Figures captions:**

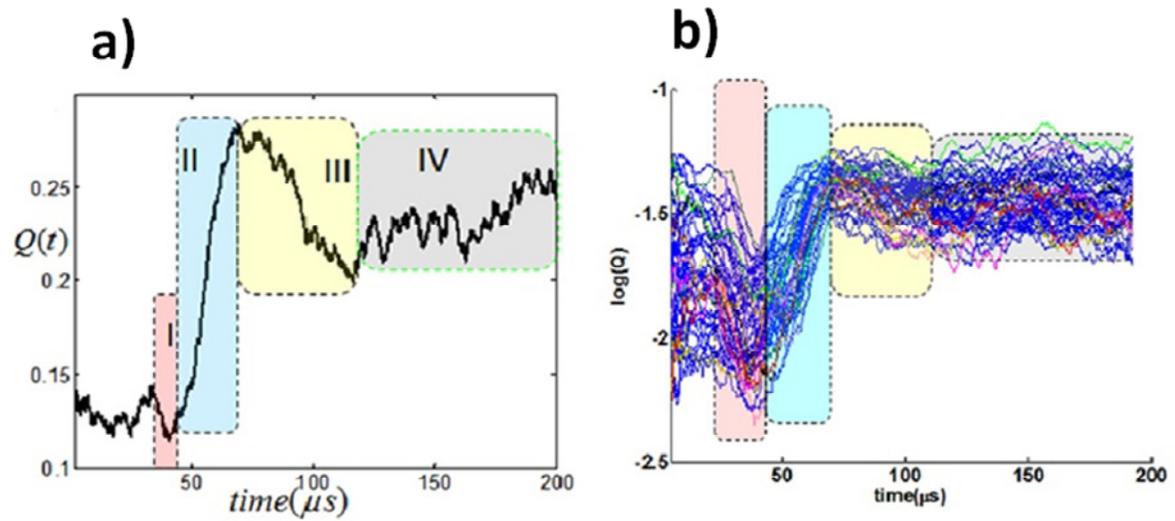

**Figure 1| General dynamic of modularity index of precursory events.**

 **a,** A typical evolution of modularity index of a precursor rupture front from LabEQ1. 4 main phases are: approaching and failure time (Phase I , $Q(t) \rightarrow Q_{min}(t)$ ), rapid growth of modularity (Phase II) , slow drop of modularity ( phase III) and slow growth (Phase IV). **b,** An approximate collapse of the modularity values for recorded regular ruptures ( i.e., ~20 events from R2 in figures 2 and 3) . The duration of phase I and II are nearly constant approximating 5-10 $\mu s$ and 20-30 $\mu s$ respectively (also see Fig.S4-S6).

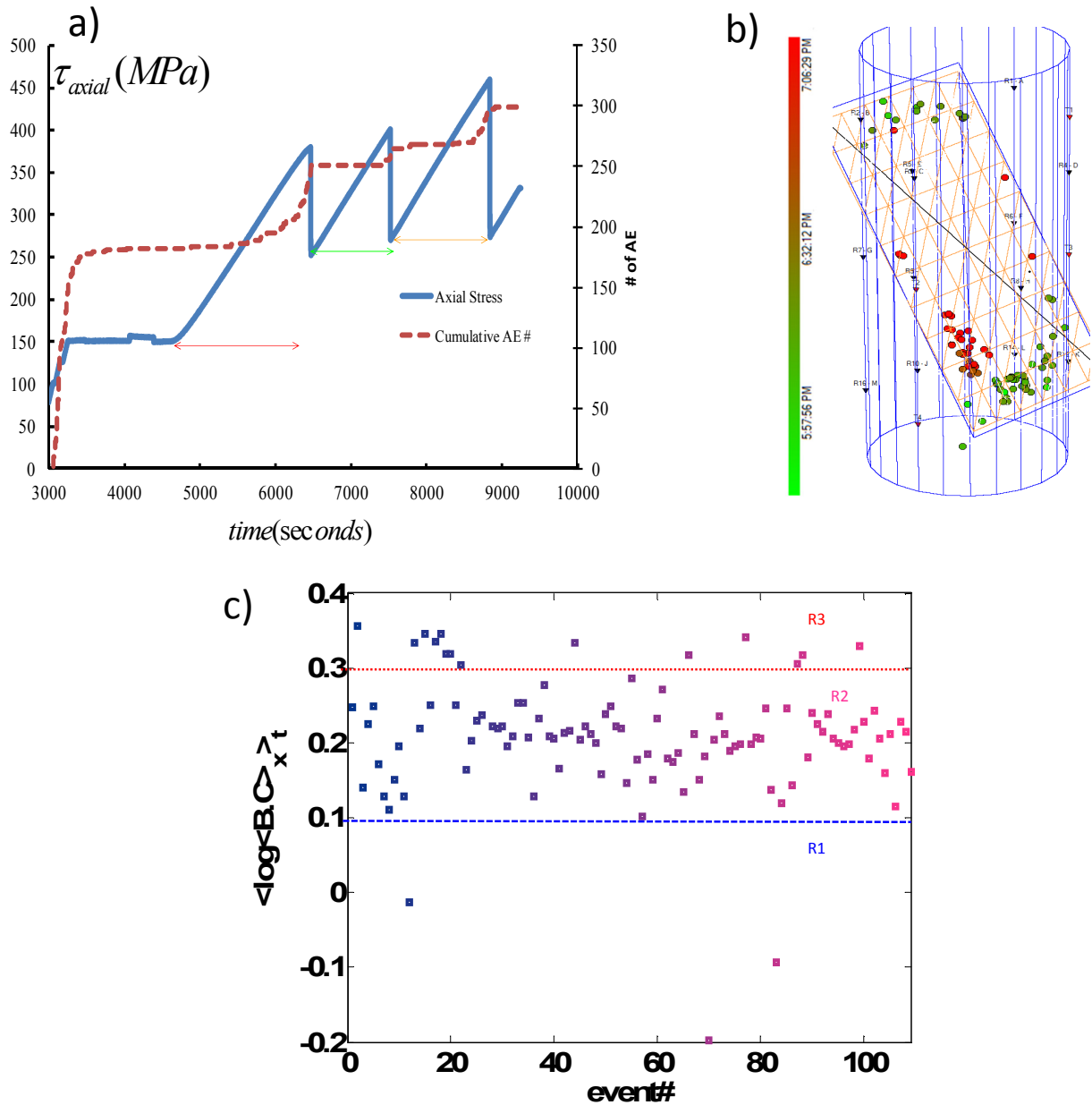

**Figure 2| Study of arrested rupture fronts with functional acoustic-friction networks reveals the evolution of acoustic events.**

**a,** Three cycles of main stick-slip events in LabEQ1 and cumulative acoustic events [31] .**b,** The geometry of the smooth fault embedded in a cylindrical westerly Granite and the source locations of the recorded events with the occurrence time of each event as the color bar. The triangles show the position of piezoelectric transducers. **c,** Mapping the recorded events in $\overline{\log <B.C>}$ versus the event's number shows some abnormal fronts (R1 and R3) . Ruptures with relatively high energy allocate lower $\overline{\log <B.C>}$ .

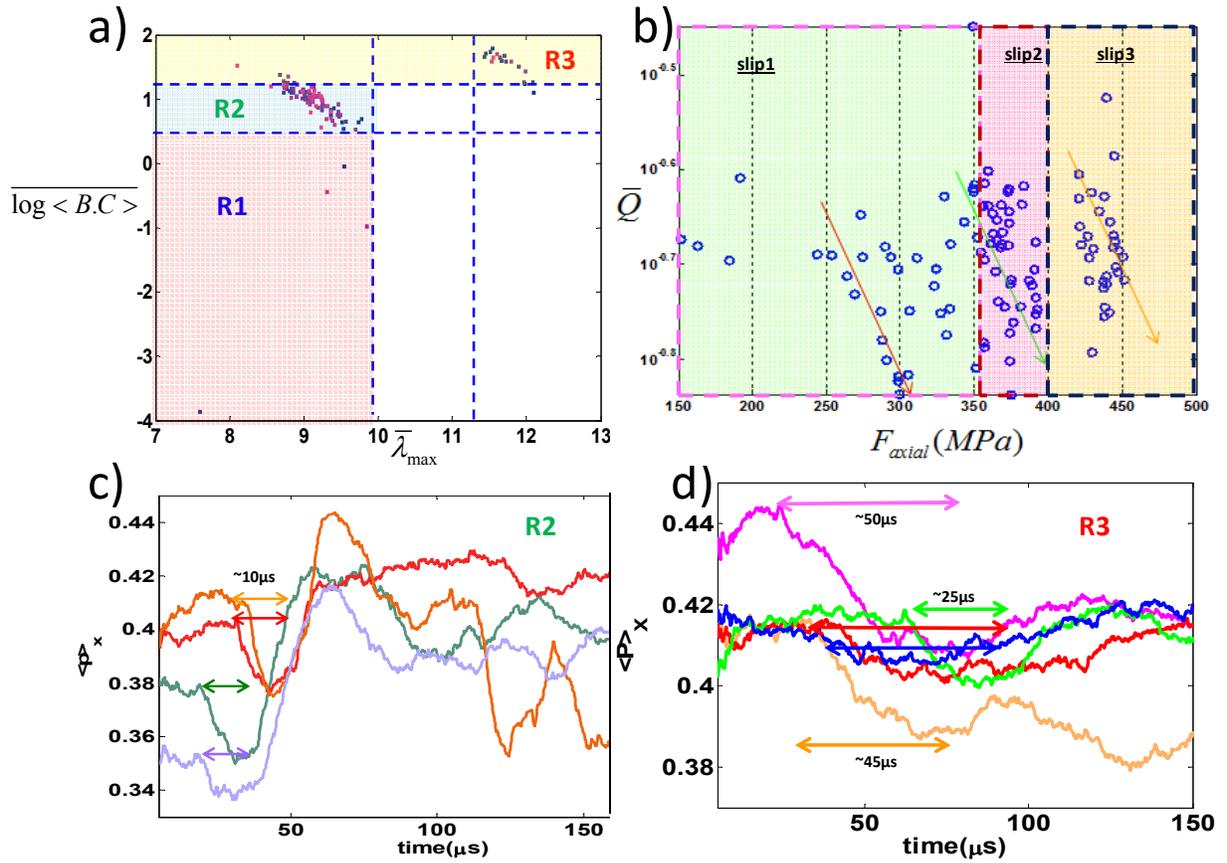

**Figure 3| Classified detachment fronts in network parameter space.**

**a,** Mapping 109 events during 204μs time-window from LabEQ1 on the spatio-temporal mean of betweenness centrality -mean of maximum eigenvalues of Laplacain parameter space. **b,** evolution of $\overline{Q}$ versus axial stress (MPa) as the remote driving stress in the smooth-fault experiment through 3 cycles of main stick-slip as has been shown in Figure 2a. A simple logarithmic relation per each cycle can be assigned as: $F_{axial} \simeq -0.33 \times 10^3 \log \overline{Q}$ indicating that generally the lower values of $\overline{Q}$ occur in relatively higher driving shear stress . **c,** a typical characteristic of the regular recorded waveforms with respect to the mean inverse participation coefficient shows ~10μs sharp drop of localization of eigenvectors (corresponding to the phase I in figure1a) . **d,** Recorded fronts in R3 cluster show longer phase I (~25-50 μs). ($\overline{...}$) and $< ... >_x$ indicate the average over 204μs per each recorded rupture front and the mean over nodes, respectively.

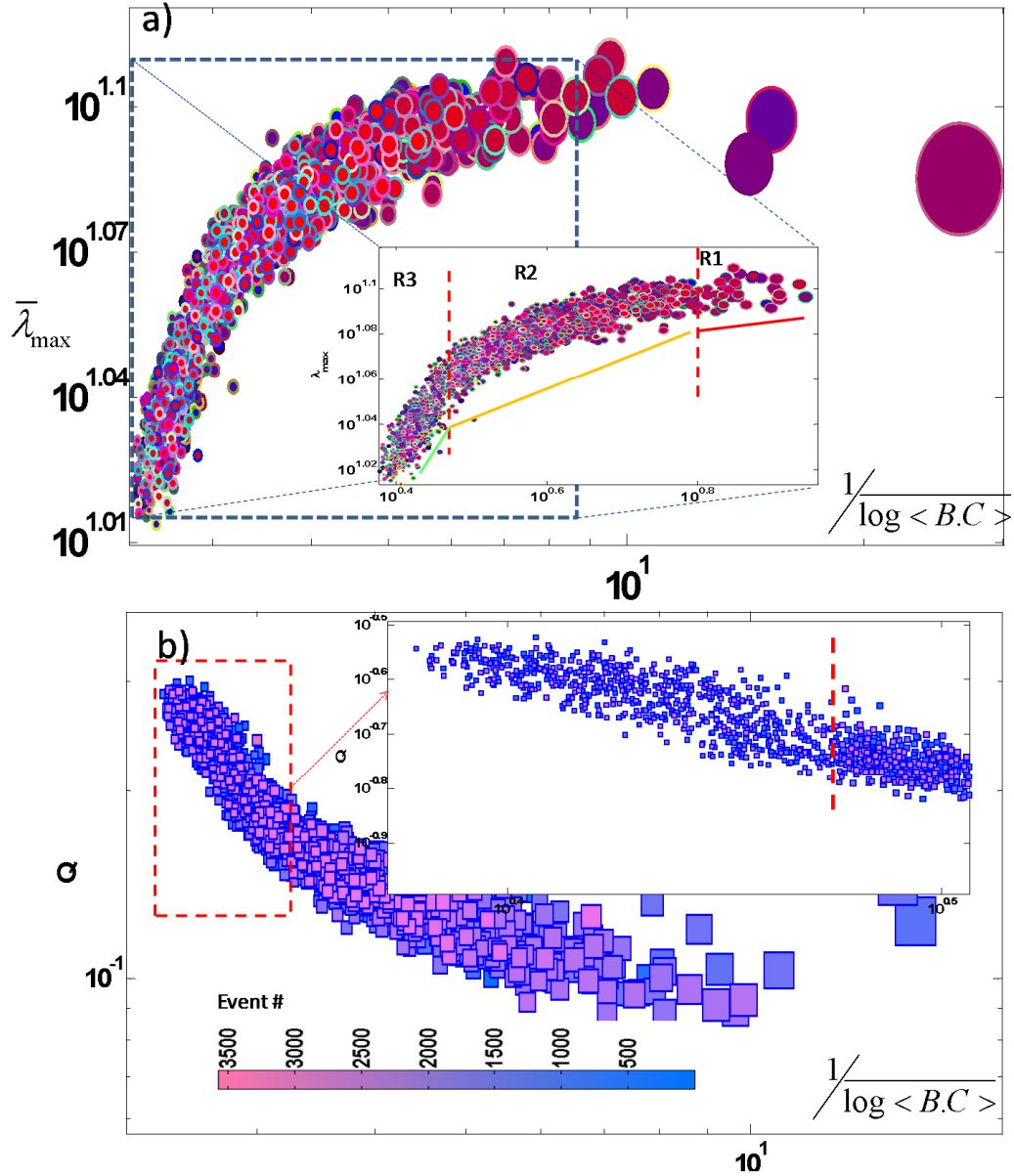

**Figure 4| Precursor events from the evolution of rough frictional interface.**

Evolution of rupture fronts in a rough fault (LabEQ2) through the first cycle of loading in **a,** $\frac{1}{\log <B.C>} - \overline{<\lambda_{max}>}$ parameter space for ~3500 recorded acoustic emission events; a critical exponent such as $\beta$ in $<\overline{\lambda_{max}}> \sim \overline{\log <B.C>}^{-\beta}$ can be defined as it has been shown in the inset . The size of the circles is corresponding to the maximum root mean square of the amplitudes of all piezoelectric sensors .**b,** $\frac{1}{\log <B.C>} - \overline{Q}$ for the same events.

Inset shows a transition to $R_2$ (*i.e.,* regular events) occurs after a threshold level (also see Fig.S6 for loading configurations and FigS7 for the second cycle of the loading).

# Acoustic-Friction Networks and the Evolution of Precursory Rupture Fronts in Laboratory Earthquakes

H.O.Ghaffari [1],R.P.Young

*Department of Civil Engineering and Lassonde Institute, University of Toronto, Toronto, ON, Canada*

---

[1] h.o.ghaffari@gmail.com;
*MB108-170 College Street, Department of Civil Engineering and Lassonde Institute, University of Toronto, Toronto, Canada M5S3E3;*

## More Features from Acoustic-Friction Networks

In this supplementary document, we present more features of the employed algorithm on the recorded acoustic waveforms. Preparation of the samples for both Lab.Eq1 and Lab.Eq2 (Fig.S.1) and the results regarding source mechanism and position has been addressed in the previous publications [5, 6].

In figure. S2 , we have shown the typical evolution of the modularity index and other parameters (log (B.C), with-in community index (z) and assortativity (r) from node 10) from Lab.Eq1. The evolutionary phases of the mentioned phases shows a remarkable correlation with the recent reported displacement measurement from PMMA samples (Fig.S2b) while the phase II –indicated with notation "c" in Fig.S2a- in westerly granite samples is shorter than the reported results for PMMA [1]. While the emergence of the phase II in PMMA was related to the duration of the heat diffusion with respect to the released heat from the deformation (thermally induced weakening) in a tiny layer, the similar model on westerly granite does not support the obtained duration. We used a 1D thermal-diffusion model (as well as [1]) , and simultaneously change the fracture energy and the thickness of the layer while the thermal diffusivity ,density ,heat capacity are assigned for Westerly granite as follows:

```
c_p =840 (J/kg.°C);
alpha=1.29*10^-6 (m²/sec);
```

For a range of the fracture energy for Westerly Granite, the results show a fast cooling phase in less than 6-10 $\mu s$ to below ~400 °c (for high fracture energy) ,a temperature in which possible plastic deformation has been reported for Granite (Fig.S2 e). Maybe the acceptable explanation for phase II and immediate drop (phase III) is fast slipping due to crushing asperity and arresting phase of the slipping by the next immediate spatial asperity, distanced by the characteristic of slip-weakening length (such as D). The phase IV is explained with the vibration of the interface, encoded in slow decaying part. Regardless of the exact mechanism, the general concept of fracture weakening, mechanism does satisfy the observation in brittle and less brittle materials (such as PMMA).

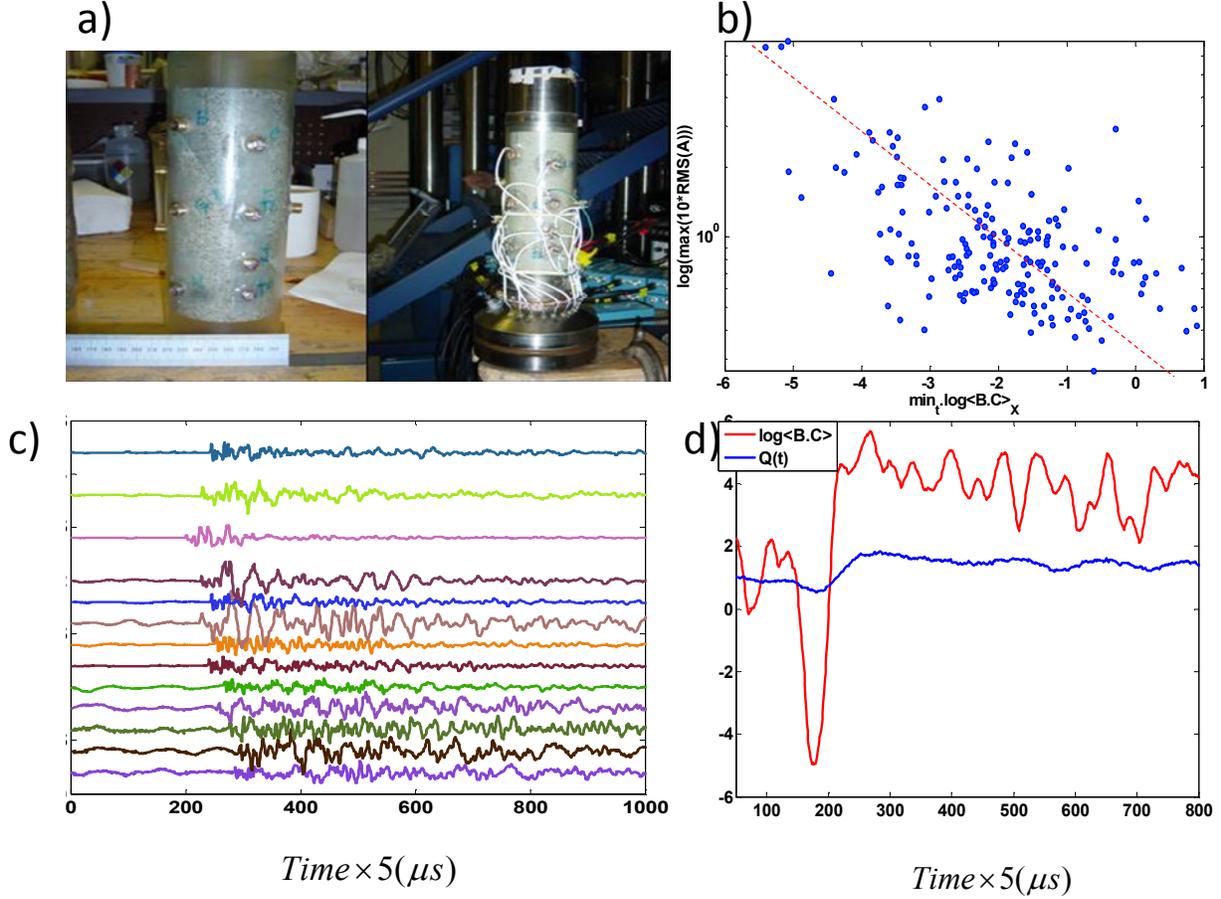

**Figure S1. a)** Natural (Left) and smooth (right) faults embedded in cylindrical Westerly granite rocks .**b)** Distribution of maximum root mean square (RMS) of waveforms for ~100 precursory rupture fronts from the first cycle of Lab.Eq2 case versus minimum of spatial mean of local flow energy in 204 μs shows a linear trend. **c, d)** a recorded event (scaled)  and its corresponding relative $\frac{Q(t)}{Q(t_0)}$ and $\frac{\log <B.C>}{\log <B.C>_0}$ .

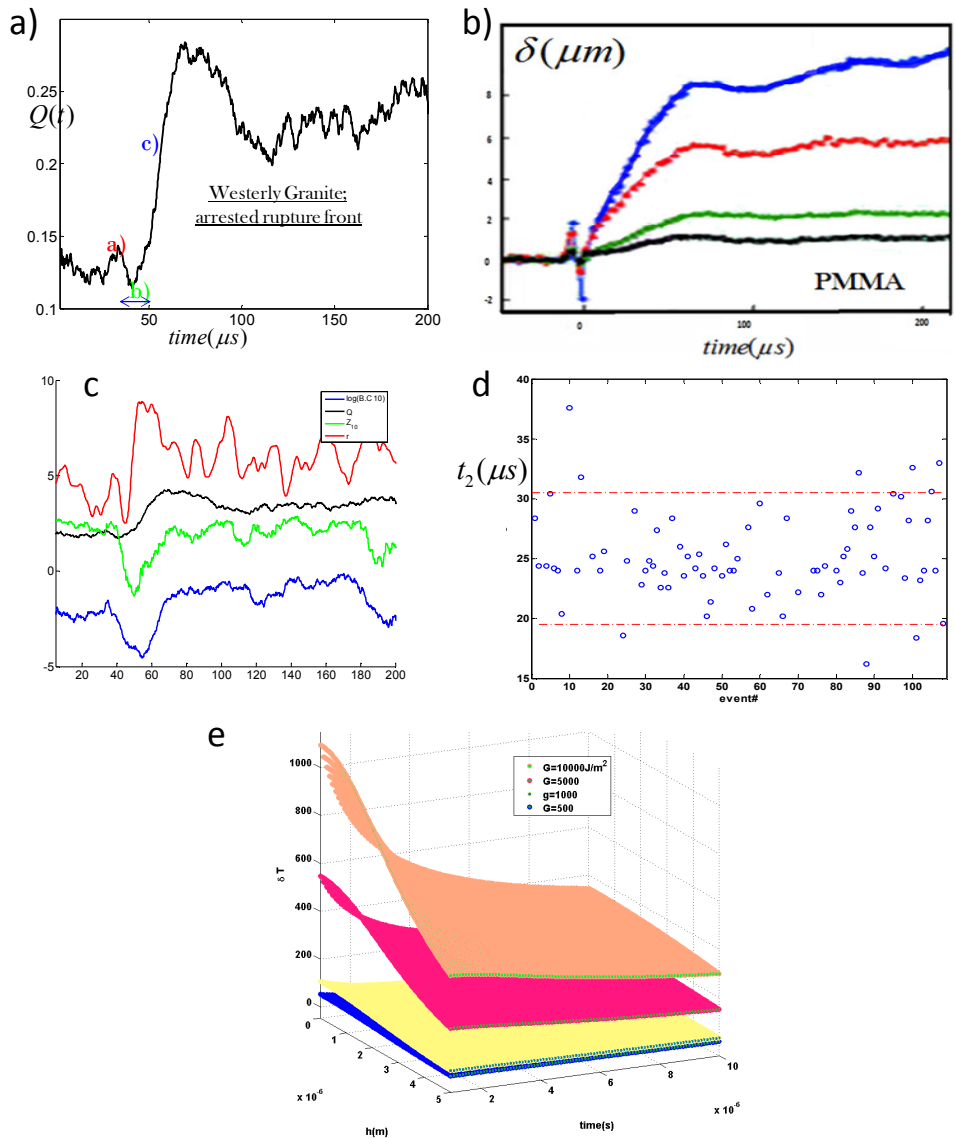

**Figure S2. a)** A typical evolution of modularity index for an precursor event LabEQ1. The three phases (a,b and c) are in agreement with the conceptual model of asperity crushing ; the approaching crack tip and crushing of an asperity are encoded in stages a and b . The rapid slip as the result of crushing the asperity occurs in stage c; we notice that the evolution of Q is the mean representative of the rupture in each "node". **b)** The measured displacement from precursor events on PMMA –frictional interface reported in [1]. The failure of asperity is imprinted with high fluctuation which is followed by a rapid slip stage. **c)** Scaled features of the functional networks from a precursory event (Lab.Eq1). The presented structural properties of the networks are: assortativity mixing (r) ,Q (modularity) , with-module index for node #10 ($Z_{10}$) and logarithmic of betweenness centrality for node #10 ($\log(B.C)_{10}$). **d)** The distribution of the duration of the phase II (phase c in part a) indicates around ~20-35 $\mu s$ . **e)** the results of 1D heat diffusion model (and possible cooling down time as phase II) with respect to 4 values o f fracture energy (from mode I to mode II) and a range of process zone (1-5 $\mu m$) in 10 $\mu s$ .

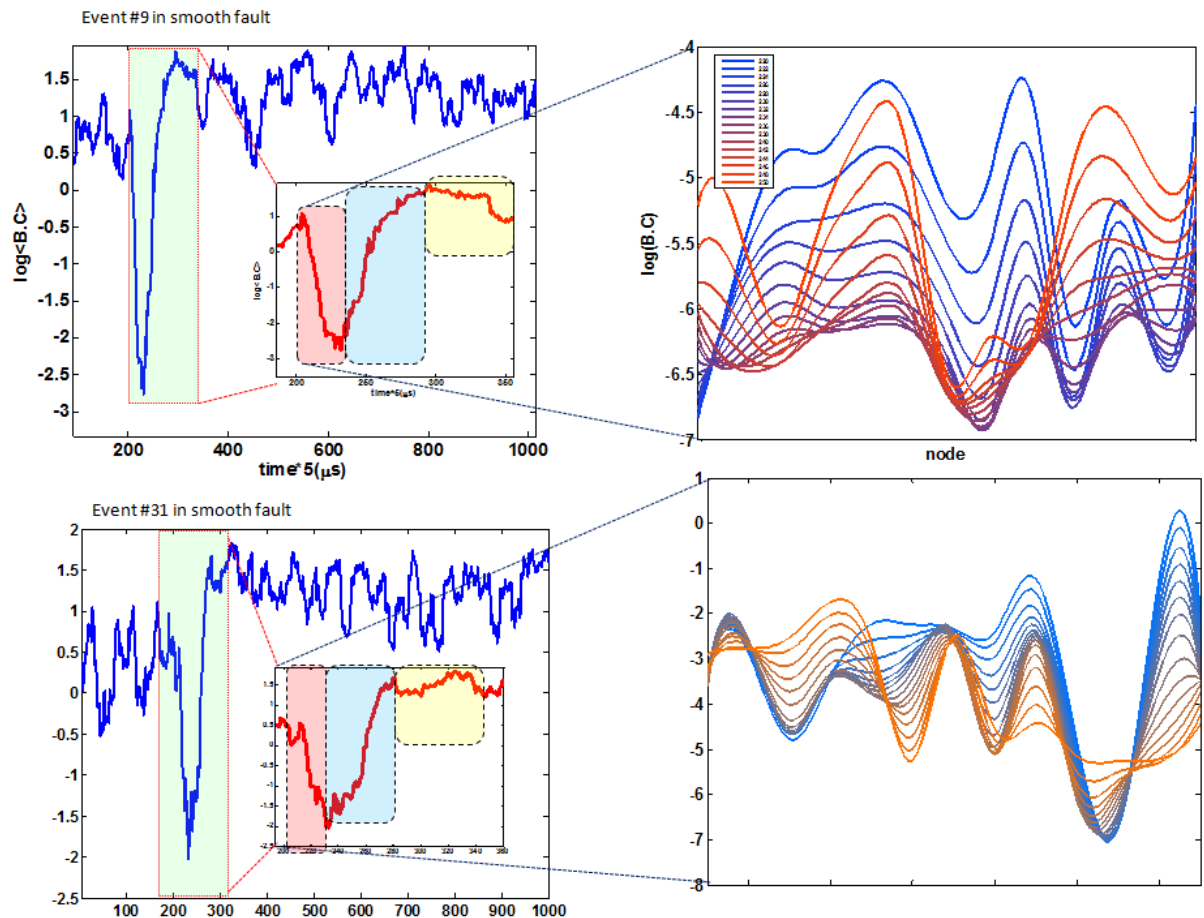

**Figure S3.** (**Left panel**) The mean-spatial evolution of betweenness centrality from two typical precursory events (#9 and #31) - LabEQ1- and (**Right panel**) the spatial evolution of "network-local energy flow" . blue color shows the dropping part (phase 1&2) and the red color show the transition to rapid slip phase. The shape and the topology of the spatial landscape of network-local energy flow (i.e., log (B.C)) does change when the betweenness centrality approaches to its minimum value. We have just shown a part of fast slip phase (Phase II).

In figure S3, we have shown the spatial variation of local flow energy index (B.C) for two arrested rupture fronts. Transition from phase I to Phase II is accompanied with the remarkable change of spatial profiles while approaching to minimum modularity (or the basin of minimum local energy flow) is followed by a fast rebounding of the profiles (where the maximum rebounding phase occurs in different basin-nodes) . We also support the idea of long period of the last phase of the evolution (phase IV) in figure S.4. Plotting typical events from Lab.EQ 1 and Lab.Eq2 regarding relative rest values shows that the slow decay phase for Lab.Eq2 events are more remarkable, indicating that the last evolutionary phase(without considering the healing process) is more affected by the topology and heterogeneity of the interface and process zones . The absolute mean rate of the decay for this phase has been shown in figure S5. A simple power law can be fitted to the collapsed data set for both cases. For Lab.EQ1, we have: $|RQ_4| \sim Q_{max}(t)^\alpha \ \alpha \approx 2.5$ . The power law coefficient is slightly higher for the Lab.Eq2, indicating a faster time characteristic for this phase.

Page | 6

Following the same idea for phase II, we find that the dependency of the rate of the growth in this phase on the maximum relative Q(t) ,with the similar power law with much smaller power coefficient ( $\alpha \approx 1.5$ ) indicating much shorter phase with the remarkable nearly constant duration.

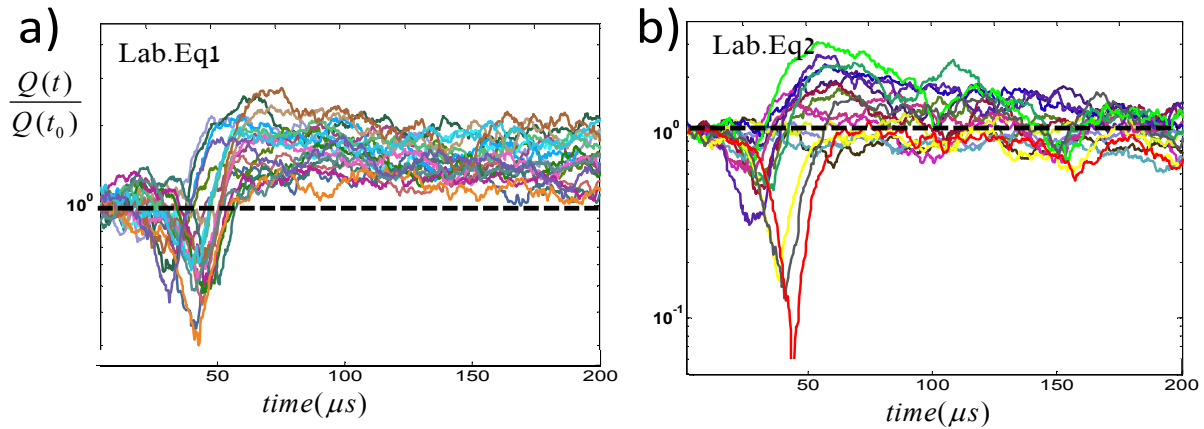

**Figure S4.** Temporal patterns of regular ruptures from Lab.Eq1 and Lab.Eq2 . **a,b)** Evolution of $\frac{Q(t)}{Q(t_0)}$ versus time for regular ruptures from Lab.Eq.1 and Lab.Eq2 . Ruptures from rough -natural fault reach in a relatively faster rate to the rest state of the parameters indicated with the dashed line.

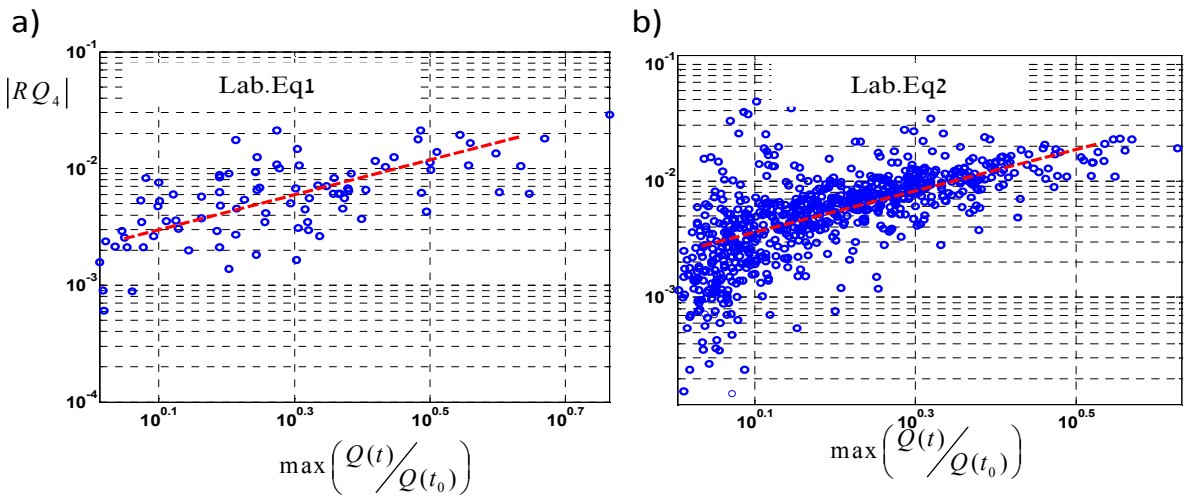

**Figure S5. a)** Absolute rate of decay for phase 4 (RQ$_4$) for ~100 rupture fronts in Lab.Eq1 versus maximum relative Q(t) . **b)** the same parameter for Lab.Eq2 over ~1400 events . Events with higher max.Q(t) decays faster than events with relatively smaller maximum modularity.

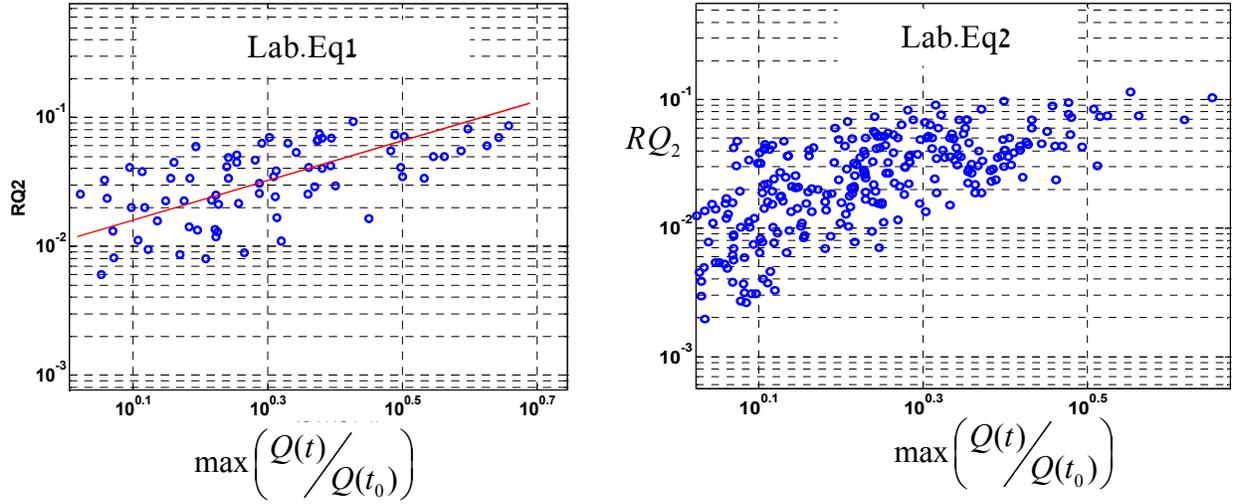

**Figure S6. a)** Rate of growth in phase II (RQ2) for ~100 rupture fronts in Lab.Eq1 versus maximum relative Q(t) .**b)** the same parameter for Lab.Eq2 over ~400 events . A simple scaling relation can be fitted as well as: $RQ_2 \sim Q_{max}(t)^\alpha \; \alpha \approx 1.5$ ,revealing a nearly constant duration of phase II.

To support the idea of the possible scaling of the temporal mean of the modularity with the induced uniform shear stress we have shown two more network-parameters (mean of maximum Laplacian of eigenvalues and spatio-temporal of the centrality) in figure S.7. Remarkable separation of some rupture fronts are the indications of relatively abnormal rupture fronts. In this case, as we have shown in the main text (Fig.3c), the most of the separated events have longer phase I, implicating a slow deformation.

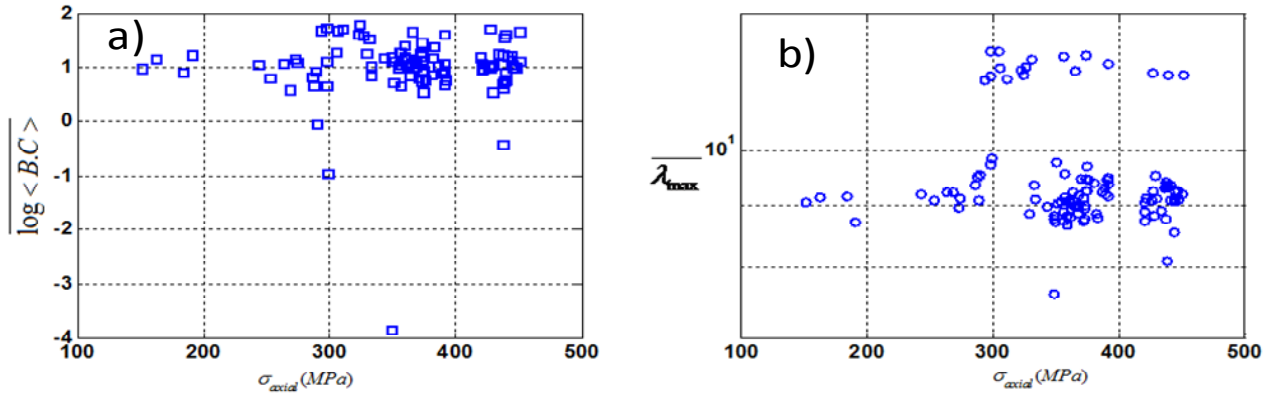

**Figure S7. a)** Evolution of $\overline{\log <B.C>}$ and **b)** $\overline{<\lambda_{max}>}$ versus axial stress (MPa) versus the remote driving stress (shear stress) in the smooth-fault experiment through 3 cycles of main stick-slip.

In figure S.8 , we have shown the loading configuration on a rough-natural fault (Lab.Eq2-also see [9]). The spatial distribution of three networks' parameters through the first cycle of the loading has been shown in Fig.S8c . The second cycle of the loading –Figure S9-is accompanied with the clipped events (and the saturation of the acquisition system) which does not allow the acquisition of information regarding the occurred events. The separation of events for high values of $\bar{Q}(t)$ is distinguished in part <u>a</u> Figure S 9, confirming our hypothesis on R3 ruptures. Then, we showed even if the nature of events and their spatial distribution in most of the cases are complex; however their signatures in proper metrics (such as networks parameter spaces) follow universal patterns, helping to understand the damage evolution in materials and eventually presenting much more precise models. Collectively, our results show that the dynamic of the networks are not random and are coupled with the dynamic of the rupture's modes (one can compare the obtained network parameters with the random (null) network models.

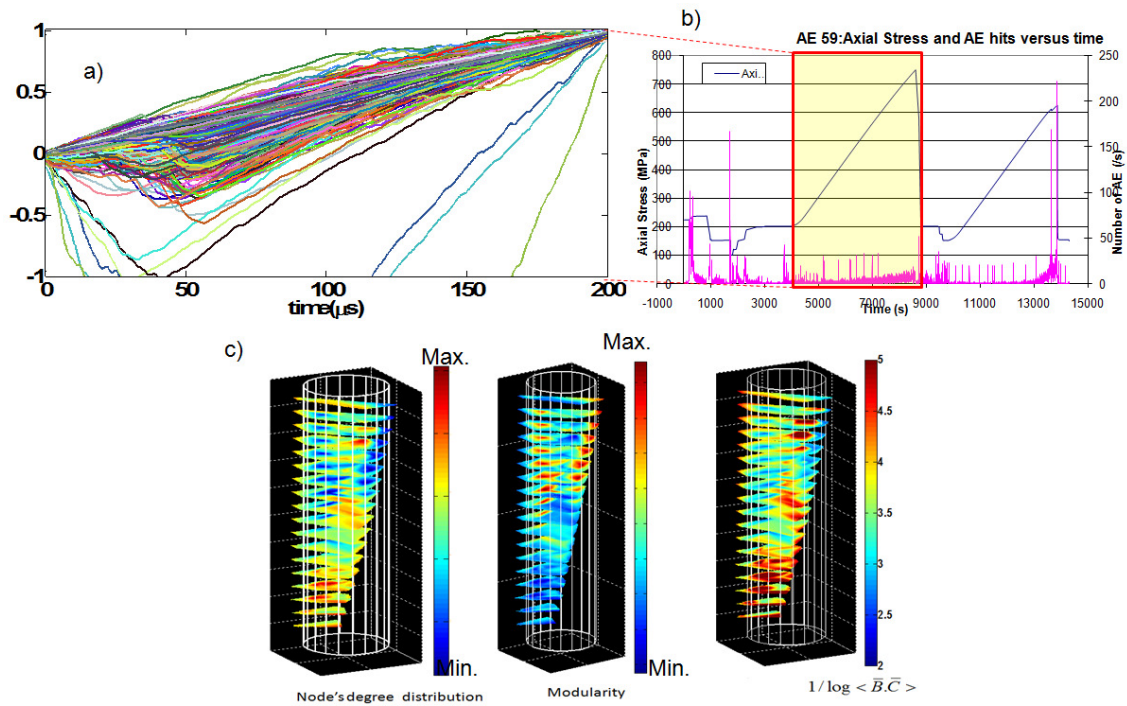

**Figure S8.** LabEQ2: First cycle of loading of the rough-fault experiment [8] **a)** variation of normalized $\delta(t) \equiv \int_0^t \log < B.C > dt$ parameter over ~3000 recorded waveforms in natural rough rock-fault; **b)** the loading condition and cumulative acoustic emission hits per second ; the red window is the first cycle; **c)** the spatial variations of node's degree, modularity and $\dfrac{1}{\log < B.C >}$ (from left to right) , respectively. To plot c, we have considered the location of events and allocate the network parameters in ~204 $\mu s$ as the monitored interval per each rupture front.

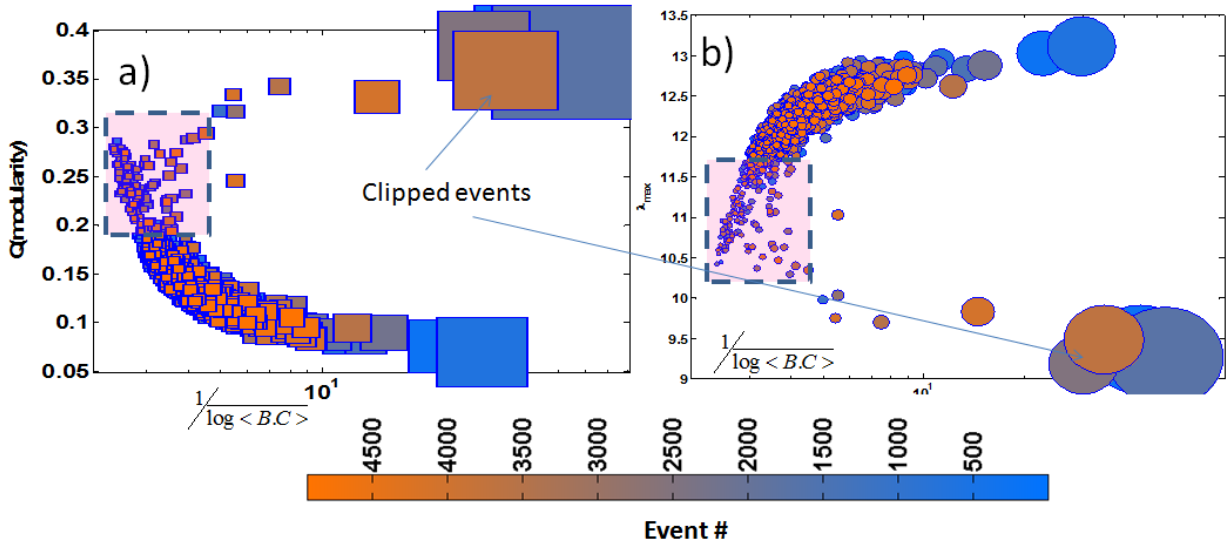

**Figure S9**. **a)** The second period of the loading (figure S8.b) in the rough-fault experiment over 4500 successive events (confining pressure was 150 MPa) ; a) $\frac{1}{\log <B.C>} - \overline{Q}$ ;the separation of events for high modular events is obvious; **b)** $\frac{1}{\log <B.C>} - \overline{\lambda_{max}}$ parameter space for the same events ; most of recorded precursor rupture fronts follow a universal pattern that can be defined with a universal exponent in $\overline{\lambda_{max}} \sim \overline{\log <B.C>}^{-\beta}$ in which high, intermediate and low values of $\beta$ correspond to $R_3, R_2$ and $R_1$ ruptures ,respectively.

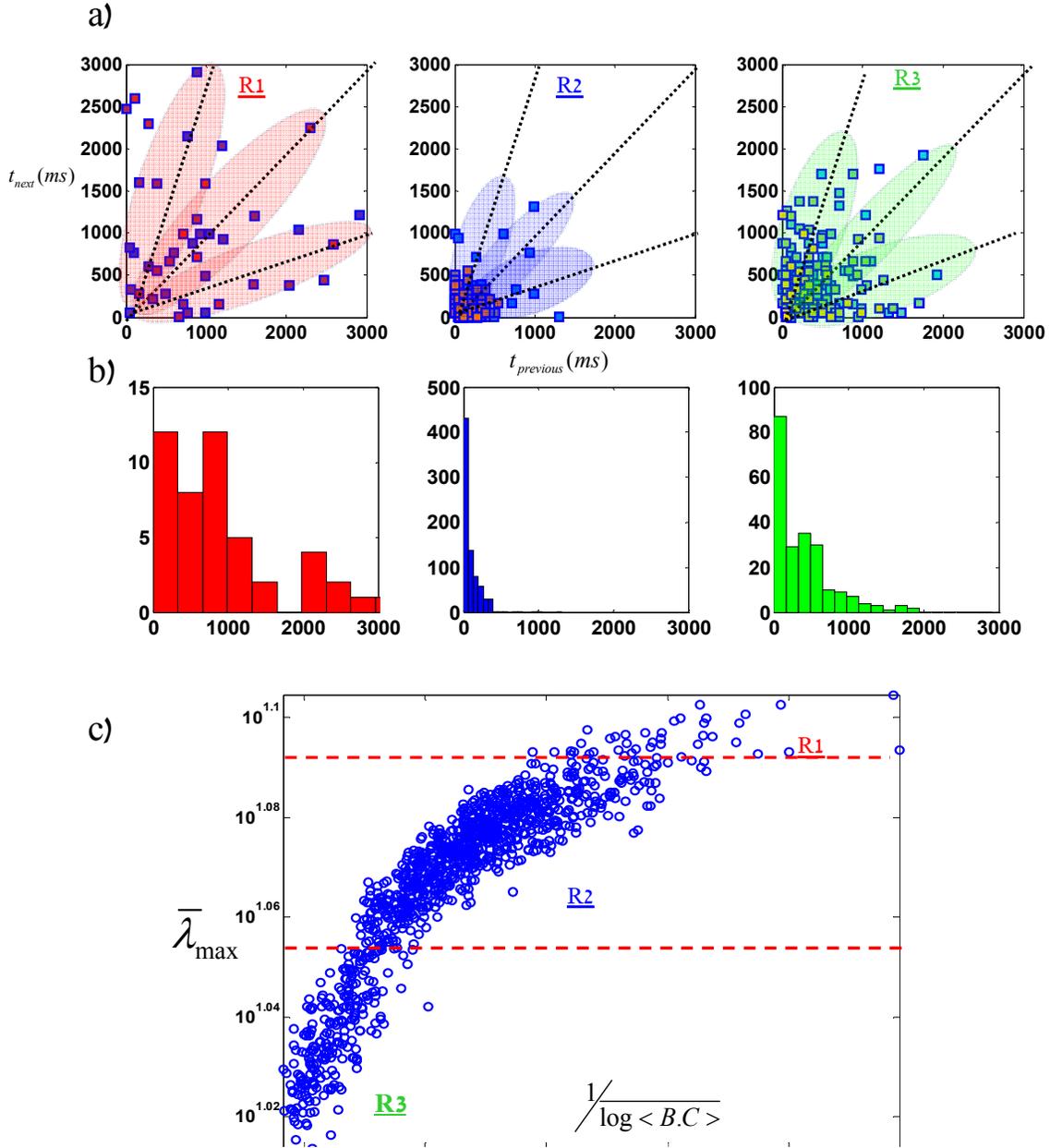

**Figure S10. a)** Temporal complexity (subsequent versus preceding waiting times ) of events classified in R1,R2 and R3 from ~1100 rupture front from the first cycle of LabEQ2 ;**b)** distribution of recurrence time for each cluster shown in **(c). ** Scaling R1 with longer recurrence time Events indicates that events with high energies (i.e., higher $\frac{1}{\log <B.C>}$ ) scale with longer recurrence times but the inverse proposition is not correct ; In other words, the events with relatively longer recurrence times not necessarily correlate with events with high released energy (R3 zone).